\documentclass[11pt]{article}
\usepackage{amsmath,amssymb,amsthm,amsxtra,overpic,bbm,bm,epsfig,ulem,color,multirow}
\usepackage{braket}

\begin{document}

\begin{center}
{\Large Leptogenesis consequences of trimaximal mixing and $\mu$-$\tau$ reflection symmetry \\
in the most minimal seesaw model}
\end{center}

\vspace{0.05cm}

\begin{center}
{\bf Zhen-hua Zhao, Hong-Yu Shi, Yan Shao\footnote{shaoyan0998@163.com}} \\
{ $^1$ Department of Physics, Liaoning Normal University, Dalian 116029, China \\
$^2$ Center for Theoretical and Experimental High Energy Physics, \\ Liaoning Normal University, Dalian 116029, China }
\end{center}

\vspace{0.2cm}

\begin{abstract}
In this paper we have studied the realizations of the popular TM1 neutrino mixing and neutrino $\mu$-$\tau$ reflection symmetry (which are well motivated from the neutrino oscillation data and lead to interesting phenomenological consequences) in the most minimal seesaw model with a pseudo-Dirac pair of right-handed neutrinos, and their consequences for leptogenesis. In order to realize the low-scale resonant leptogenesis scenario, we have considered two possible ways of generating the tiny mass splitting between the two right-handed neutrinos: one way is to modify their Majorana mass matrix to a form as shown in Eq.~(\ref{2.2.3}); the other way is to consider the renormalization-group corrections for their masses. For the $\mu$-$\tau$ reflection symmetry, in order for leptogenesis to work, we have further considered the flavor-dependent conversion efficiencies from the lepton asymmetry to the baryon asymmetry during the sphaleron processes, and its breaking via the renormalization-group corrections.
\end{abstract}

\newpage

\section{Introduction}

As we know, the phenomena of neutrino oscillations indicate that at least two of the three neutrinos are massive (while the lightest one is still allowed to be exactly massless) and their flavor eigenstates $\nu^{}_\alpha$ (for $\alpha =e, \mu, \tau$) are certain superpositions of the mass eigenstates $\nu^{}_i$ (for $i =1, 2, 3$) possessing definite masses $m^{}_i$: $\nu^{}_\alpha = \sum^{}_i U^{}_{\alpha i} \nu^{}_i$ with $U^{}_{\alpha i}$ being the $\alpha i$ element of the $3 \times 3$  neutrino mixing matrix $U$ \cite{xing}. In the standard parametrization, $U$ is expressed in terms of three mixing angles $\theta^{}_{ij}$ (for $ij=12, 13, 23$), one Dirac CP phase $\delta$ and two Majorana CP phases $\rho$ and $\sigma$ as
\begin{eqnarray}
U  = \left( \begin{matrix}
c^{}_{12} c^{}_{13} & s^{}_{12} c^{}_{13} & s^{}_{13} e^{-{\rm i} \delta} \cr
-s^{}_{12} c^{}_{23} - c^{}_{12} s^{}_{23} s^{}_{13} e^{{\rm i} \delta}
& c^{}_{12} c^{}_{23} - s^{}_{12} s^{}_{23} s^{}_{13} e^{{\rm i} \delta}  & s^{}_{23} c^{}_{13} \cr
s^{}_{12} s^{}_{23} - c^{}_{12} c^{}_{23} s^{}_{13} e^{{\rm i} \delta}
& -c^{}_{12} s^{}_{23} - s^{}_{12} c^{}_{23} s^{}_{13} e^{{\rm i} \delta} & c^{}_{23}c^{}_{13}
\end{matrix} \right) \left( \begin{matrix}
e^{{\rm i}\rho} &  & \cr
& e^{{\rm i}\sigma}  & \cr
&  & 1
\end{matrix} \right) \;,
\label{1}
\end{eqnarray}
where the abbreviations $c^{}_{ij} = \cos \theta^{}_{ij}$ and $s^{}_{ij} = \sin \theta^{}_{ij}$ have been employed.

Thanks to the various neutrino oscillation experiments, three neutrino mixing angles and the neutrino mass squared differences $\Delta m^2_{ij} \equiv m^2_i - m^2_j$ have been measured to a good degree of accuracy, and there is also a preliminary result for $\delta$ but with a large uncertainty. Several research groups have performed global analyses of the accumulated neutrino oscillation data to extract the values of these parameters \cite{global,global2}. For definiteness, we will use the results in Ref.~\cite{global} (reproduced in Table~1 here) as reference values in the following numerical calculations. Note that the sign of $\Delta m^2_{31}$ remains undetermined, thereby allowing for two possible neutrino mass orderings: the normal ordering (NO) $m^{}_1 < m^{}_2 < m^{}_3$ and inverted ordering (IO) $m^{}_3 < m^{}_1 < m^{}_2$. Unfortunately, neutrino oscillations are completely insensitive to the absolute scales of neutrino masses and the Majorana CP phases. Their values can only be inferred from certain non-oscillatory experiments such as the neutrinoless double beta decay experiments \cite{0nbb}. But so far there has not been any lower bound on the lightest neutrino mass (allowing it to be exactly massless), nor any constraint on the Majorana CP phases.

From Table~1 one can see that $\theta^{}_{12}$ and $\theta^{}_{23}$ are close to some special values (i.e., $\sin^2 \theta^{}_{12} \sim 1/3$ and $\sin^2 \theta^{}_{23} \sim 1/2$). And $\theta^{}_{13}$ is relatively small. In fact, before its value was measured, it had been widely expected to be vanishingly small. For the ideal case of $\sin^2 \theta^{}_{12} = 1/3$, $\sin^2 \theta^{}_{23} = 1/2$ and $\theta^{}_{13} =0$, the neutrino mixing matrix takes a very special form as
\begin{eqnarray}
U^{}_{\rm TBM}= \displaystyle \frac{1}{\sqrt 6} \left( \begin{array}{ccc}
2 & \sqrt{2} & 0 \cr
1 & - \sqrt{2}  & -\sqrt{3}  \cr
1 & - \sqrt{2}  & \sqrt{3} \cr
\end{array} \right)  \;,
\label{2}
\end{eqnarray}
which is referred to as the tribimaximal (TBM) mixing \cite{TB}. Such a remarkably simple and compact neutrino mixing pattern has attracted people to believe that there exists some flavor symmetry in the neutrino sector and have made a lot of attempts to explore the possible flavor symmetries underlying the observed neutrino mixing pattern \cite{FS}. In the literature, there are two mainstream approaches for the flavor-symmetry studies: one approach is to arrange a certain flavor symmetry in the lepton sector and then break it in different ways in the charged-lepton and neutrino sectors so that a non-trivial neutrino mixing pattern may arise; the other approach is to explore the minimal flavor symmetries exhibited by the neutrino flavor parameters. In the light of the experimental results for the neutrino flavor parameters, two very attractive candidates of the latter approach are the trimaximal mixing and $\mu$-$\tau$ reflection symmetry, which are briefly introduced as follows.

On one hand, the observation of a relatively large $\theta^{}_{13}$ (compared to $0$) compels us to make corrections for the TBM mixing. A very popular and minimal choice is to retain its first or second column while modifying the other two columns, giving the first or second trimaximal (TM1 or TM2) mixing \cite{TM}
\begin{eqnarray}
U^{}_{\rm TM1}=  \displaystyle \frac{1}{\sqrt 6} \left( \begin{array}{ccc}
2 & \cdot & \cdot \cr
1 & \cdot & \cdot \cr
1 & \cdot & \cdot \cr
\end{array} \right)  \;, \hspace{1cm}
U^{}_{\rm TM2}=  \displaystyle \frac{1}{\sqrt 3} \left( \begin{array}{ccc}
\cdot & 1 & \cdot \cr
\cdot & -1 &  \cdot \cr
\cdot &  -1 & \cdot \cr
\end{array} \right)  \;.
\label{3}
\end{eqnarray}
Due to its simplicity and interesting consequences, the trimaximal mixing has attracted a lot of attention and many flavor symmetries have been employed to build realistic neutrino mass models realizing it \cite{FS}.

On the other hand, owing to the preliminary experimental hint for $\delta \sim - \pi/2$ \cite{T2K}, the $\mu$-$\tau$ reflection symmetry \cite{mu-tauR, mutau2} has become increasingly popular, under which the neutrino mass matrix is required to keep invariant with respect to the following transformations of three left-handed neutrino fields
\begin{eqnarray}
\nu^{}_{e} \leftrightarrow \nu^{c}_e \;, \hspace{1cm} \nu^{}_{\mu} \leftrightarrow \nu^{c}_{\tau} \;,
\hspace{1cm} \nu^{}_{\tau} \leftrightarrow \nu^{c}_{\mu} \;,
\label{4}
\end{eqnarray}
with the superscript $c$ denoting the charge conjugation of relevant fields.
Under this symmetry, the Majorana neutrino mass matrix takes a form as
\begin{eqnarray}
M^{}_\nu = \left( \begin{matrix}
A & B & B^* \cr
B & C & D \cr
B^* & D & C^*
\end{matrix} \right) \;,
\label{5}
\end{eqnarray}
with $A$ and $D$ being real,
which leads to the following interesting predictions for the neutrino mixing angles and CP phases
\begin{eqnarray}
\theta^{}_{23} = \frac{\pi}{4} \;, \hspace{1cm} \delta = \pm \frac{\pi}{2} \;,
\hspace{1cm} \rho = 0 \ {\rm or} \ \frac{\pi}{2} \;, \hspace{1cm} \sigma = 0 \ {\rm or} \ \frac{\pi}{2} \;.
\label{6}
\end{eqnarray}

As for the neutrino masses, one of the most popular and natural ways of generating them is the type-I seesaw model in which at least two heavy right-handed neutrinos $N^{}_I$ ($I=1, 2, ...$) are introduced into the Standard Model (SM) \cite{seesaw}. First of all, $N^{}_I$ can constitute the Yukawa coupling operators together with the left-handed neutrinos $\nu^{}_\alpha$ (which reside in the lepton doublets $L^{}_\alpha$) and the Higgs doublet $H$: $(Y^{}_{\nu})^{}_{\alpha I} \overline {L^{}_\alpha} H N^{}_I $ with $(Y^{}_{\nu})^{}_{\alpha I}$ being the $\alpha I$ element of the Yukawa coupling matrix $Y^{}_{\nu}$. These operators will generate the Dirac neutrino masses $(M^{}_{\rm D})^{}_{\alpha I}= (Y^{}_{\nu})^{}_{\alpha I} v$ [here $(M^{}_{\rm D})^{}_{\alpha I}$ is the $\alpha I$ element of the Dirac neutrino mass matrix $M^{}_{\rm D}$] after the neutral component of $H$ acquires a nonzero vacuum expectation value (VEV) $v = 174$ GeV. Furthermore, $N^{}_I$ themselves can also have the Majorana mass terms $\overline{N^c_I} (M^{}_{\rm R})^{}_{IJ} N^{}_J$ [here $(M^{}_{\rm R})^{}_{IJ}$ is the $IJ$ element of the right-handed neutrino mass matrix $M^{}_{\rm R}$]. Then, under the seesaw condition $M^{}_{\rm R} \gg M^{}_{\rm D}$, one will obtain an effective Majorana mass matrix for three light neutrinos as
\begin{eqnarray}
M^{}_{\nu} = - M^{}_{\rm D} M^{-1}_{\rm R} M^{T}_{\rm D} \;,
\label{7}
\end{eqnarray}
by integrating the right-handed neutrinos out. Thanks to such a formula, the smallness of neutrino masses can be naturally explained by the heaviness of right-handed neutrinos [e.g., ${\cal O}(10^{13})$ GeV].

Remarkably, the seesaw model also provides an attractive explanation (which is known as the leptogenesis mechanism \cite{leptogenesis, Lreview}) for the baryon-antibaryon asymmetry of the Universe \cite{planck}
\begin{eqnarray}
Y^{}_{\rm B} \equiv \frac{n^{}_{\rm B}-n^{}_{\rm \bar B}}{s} \simeq (8.69 \pm 0.04) \times 10^{-11}  \;,
\label{8}
\end{eqnarray}
where $n^{}_{\rm B}$ ($n^{}_{\rm \bar B}$) denotes the baryon (antibaryon) number density and $s$ the entropy density. The leptogenesis mechanism works in a way as follows: a lepton-antilepton asymmetry is firstly generated from the out-of-equilibrium and CP-violating decays of right-handed neutrinos and then partly converted into the baryon-antibaryon asymmetry via the sphaleron processes. It is well known that in the scenario of the right-handed neutrino masses being hierarchial, in order to successfully reproduce the observed value of $Y^{}_{\rm B}$, there exists a lower bound about $10^9$ GeV for the right-handed neutrino mass scale \cite{DI}. However, for low-scale seesaw models, a successful leptogenesis can still be achieved with the help of the resonant leptogenesis scenario which is realized for nearly degenerate right-handed neutrinos \cite{resonant}.

\begin{table}\centering
  \begin{footnotesize}
    \begin{tabular}{c|cc|cc}
     \hline\hline
      & \multicolumn{2}{c|}{Normal Ordering}
      & \multicolumn{2}{c}{Inverted Ordering }
      \\
      \cline{2-5}
      & bf $\pm 1\sigma$ & $3\sigma$ range
      & bf $\pm 1\sigma$ & $3\sigma$ range
      \\
      \cline{1-5}
      \rule{0pt}{4mm}\ignorespaces
       $\sin^2\theta^{}_{12}$
      & $0.303_{-0.012}^{+0.012}$ & $0.270 \to 0.341$
      & $0.303_{-0.012}^{+0.012}$ & $0.270 \to 0.341$
      \\[1mm]
       $\sin^2\theta^{}_{23}$
      & $0.451_{-0.016}^{+0.019}$ & $0.408 \to 0.603$
      & $0.569_{-0.021}^{+0.016}$ & $0.412 \to 0.613$
      \\[1mm]
       $\sin^2\theta^{}_{13}$
      & $0.02225_{-0.00059}^{+0.00056}$ & $0.02052 \to 0.02398$
      & $0.02223_{-0.00058}^{+0.00058}$ & $0.02048 \to 0.02416$
      \\[1mm]
       $\delta/\pi$
      & $1.29_{-0.14}^{+0.20}$ & $0.80 \to 1.94$
      & $1.53_{-0.16}^{+0.12}$ & $1.08\to 1.91$
      \\[3mm]
       $\Delta m^2_{21}/(10^{-5}~{\rm eV}^2)$
      & $7.41_{-0.20}^{+0.21}$ & $6.82 \to 8.03$
      & $7.41_{-0.20}^{+0.21}$ & $6.82 \to 8.03$
      \\[3mm]
       $|\Delta m^2_{31}|/(10^{-3}~{\rm eV}^2)$
      & $2.507_{-0.027}^{+0.026}$ & $2.427 \to 2.590$
      & $2.412_{-0.025}^{+0.028}$ & $2.332 \to 2.496$
      \\[2mm]
      \hline\hline
    \end{tabular}
  \end{footnotesize}
  \caption{The best-fit values, 1$\sigma$ errors and 3$\sigma$ ranges of six neutrino
oscillation parameters extracted from a global analysis of the existing
neutrino oscillation data \cite{global}. }
\end{table}

In this paper, following the Simplicity Principle, we will consider the possibility that there only exists two right-handed neutrinos $N^{}_1$ and $N^{}_2$, which is referred to as the minimal seesaw model \cite{MSS, MSS2}, and their Majorana mass matrix takes a form as
\begin{eqnarray}
M^{}_{\rm R}= \displaystyle \left( \begin{array}{cc}
0 & M \cr
M & 0
\end{array} \right)  \;.
\label{9}
\end{eqnarray}
Such a right-handed neutrino mass matrix can be naturally realized in the minimal linear seesaw model \cite{mLSS, LSS} and is the most minimal one in the sense that it only contains a single mass parameter.
Due to its simplicity, this form of right-handed neutrino mass matrix has received considerable attention \cite{mostMSS}. More interestingly, it can also serve as a unique starting point for the realization of low-energy seesaw and leptogenesis regimes: initially, the two right-handed neutrinos have equal masses $M$ but opposite parities [i.e., two eigenvalues of $M^{}_{\rm R}$ in Eq.~(\ref{9}) are $-M$ and $M$, respectively], forming a Dirac pair; if they acquire a tiny mass splitting through some way (thus forming a pseudo-Dirac pair \cite{pseudo}), then the resonant leptogenesis scenario will be naturally realized. In this scenario, a successful leptogenesis can be achieved even if the right-handed neutrino mass is lowered to the TeV scale, which has the potential to be directly accessed by presently running and forseeable collider experiments \cite{HNL}.

Motivated by the above facts, in this paper we will study the realizations and consequences of the above-mentioned TM1 mixing and $\mu$-$\tau$ reflection symmetry (which serve as powerful guiding principles for the organization of the flavor structure of the seesaw models) in the minimal seesaw model with $M^{}_{\rm R}$ in Eq.~(\ref{9}) and TeV-scale right-handed neutrino masses (which have the potential to be directly accessed by presently running and forseeable collider experiments). As we will see, such a combination will further reduce the parameters of the seesaw model and consequently make it more predictive.

\section{Realization and consequence of the TM1 mixing}

In this section, we study the realization and consequence of the TM1 mixing in the minimal seesaw model with $M^{}_{\rm R}$ in Eq.~(\ref{9}).

\subsection{Realization of the TM1 mixing}

We first consider the realization of the TM1 mixing in the minimal seesaw model with $M^{}_{\rm R}$ in Eq.~(\ref{9}). In a general minimal seesaw model, $M^{}_{\rm D}$ can be parameterized as
\begin{eqnarray}
M^{}_{\rm D}= \displaystyle \left( \begin{array}{cc}
a & a^\prime \cr
b & b^\prime \cr
c & c^\prime
\end{array} \right) \;,
\label{2.1.1}
\end{eqnarray}
with all the parameters being complex.
For $M^{}_{\rm R}$ in Eq.~(\ref{9}), the seesaw formula gives an $M^{}_\nu$ as
\begin{eqnarray}
M^{}_\nu =  \displaystyle - \frac{1}{M} \left( \begin{array}{ccc}
2 a a^\prime & a b^\prime + b a^\prime & a c^\prime + c a^\prime \cr
a b^\prime + b a^\prime & 2 b b^\prime  & b c^\prime + c b^\prime  \cr
a c^\prime + c a^\prime & b c^\prime + c b^\prime  & 2 c c^\prime \cr
\end{array} \right)  \;.
\label{2.1.2}
\end{eqnarray}
In order for this $M^{}_\nu$ to yield the TM1 mixing, the following relation must hold:
\begin{eqnarray}
M^{}_\nu = R^{}_{\rm TM1} M^{}_\nu R^{}_{\rm TM1}
\hspace{0.5cm} {\rm with} \hspace{0.5cm}
R^{}_{\rm TM1}= - \frac{1}{3} \left( \begin{array}{ccc}
1 & 2 & 2 \cr
2 & -2 & 1 \cr
2 & 1 & -2 \cr
\end{array} \right)  \;,
\label{2.1.3}
\end{eqnarray}
which leads to the requirements
\begin{eqnarray}
b+c = -2 a \;, \hspace{1cm} b^\prime + c^\prime = - 2a^\prime \;.
\label{2.1.4}
\end{eqnarray}
Therefore, in the minimal seesaw model with $M^{}_{\rm R}$ in Eq.~(\ref{9}), the form of $M^{}_{\rm D}$ that naturally realizes the TM1 mixing can be expressed as
\begin{eqnarray}
M^{}_{\rm D}= \displaystyle \left( \begin{array}{cc}
a & a^\prime \cr
-a(1+x) & -a^\prime(1+y) \cr
-a(1-x) & -a^\prime(1-y)
\end{array} \right) \;,
\label{2.1.5}
\end{eqnarray}
with $x$ and $y$ being two dimensionless complex parameters.

The form of $M^{}_{\rm D}$ in Eq.~(\ref{2.1.5}) can be easily realized by slightly modifying the flavor-symmetry models for realizing the TBM mixing (see, e.g., Ref.~\cite{zhao}): in the usual minimal seesaw model, under the employed flavor symmetry (e.g., the $A^{}_4$ or $S^{}_4$ group \cite{FS}), three lepton doublets are organized into a triplet representation $L=(L^{}_e, L^{}_\mu, L^{}_\tau)$ while the two right-handed neutrinos are simply singlets. To break the flavor symmetry in a proper way, two flavon fields $\phi^{}_{J}$ (for $J=1, 2$) are introduced. Each of them is a triplet (with three components $\phi^{}_{J} = [(\phi^{}_{J})^{}_1, (\phi^{}_{J})^{}_2, (\phi^{}_{J})^{}_3]^T$) under the flavor symmetry.
Owing to such a setting, the following dimension-5 operators
\begin{eqnarray}
\sum^{}_{I, J} \frac{ y^{}_{IJ}}{\Lambda}[ \overline L^{}_e (\phi^{}_{J})^{}_1 + \overline L^{}_\mu (\phi^{}_{J})^{}_2 + \overline L^{}_\tau (\phi^{}_{J})^{}_3 ] H N^{}_I \;,
\label{2.1.6}
\end{eqnarray}
will serve to generate the Dirac neutrino mass terms after the electroweak and flavor symmetries are respectively broken by non-zero VEVs of the Higgs and flavon fields. Here $y^{}_{IJ}$ are dimensionless coefficients and $\Lambda$ is the energy scale for the flavor-symmetry physics.
Then, in the mass basis of two right-handed neutrinos, the TBM mixing will naturally arise under the following two conditions: $\phi^{}_1$ only couples with $N^{}_1$ while $\phi^{}_2$ only couples with $N^{}_2$ (i.e., $y^{}_{IJ} = 0$ for $I \neq J$), which can be fulfilled by invoking an auxiliary flavor symmetry; the VEV alignments of $\phi^{}_1$ and $\phi^{}_2$ come out as
\begin{eqnarray}
\langle \phi^{}_1 \rangle \propto (1, -1, -1)^T \;, \hspace{1cm} \langle \phi^{}_2 \rangle \propto (0, -1, 1)^T \;.
\label{2.1.7}
\end{eqnarray}
Note that these two forms of flavon VEV alignments are just those that are widely invoked to realize the TBM mixing with the help of an $A^{}_4$ or $S^{}_4$ flavor symmetry. For their realization under the flavor symmetry, see Refs.~\cite{FS, CK}.
By slightly modifying the above flavor-symmetry models, one can naturally realize the form of $M^{}_{\rm D}$ in Eq.~(\ref{2.1.5}): both $\phi^{}_1$ and $\phi^{}_2$ couple with $N^{}_1$ and $N^{}_2$ simultaneously, and their VEV alignments remain as in Eq.~(\ref{2.1.7}).
This can be achieved by simply discarding the aforementioned auxiliary flavor symmetry.

Now, for $M^{}_{\rm R}$ in Eq.~(\ref{9}) and $M^{}_{\rm D}$ in Eq.~(\ref{2.1.5}), the resultant $M^{}_\nu$ appears as
\begin{eqnarray}
M^{}_\nu =  \displaystyle - \frac{a a^\prime}{M} \left( \begin{array}{ccc}
2 & - 2-x-y & - 2+x+y \cr
- 2-x-y & 2 (1+x)(1+y)  & 2(1 - x y)  \cr
- 2+x+y & 2(1 - x y)  & 2 (1-x)(1-y) \cr
\end{array} \right)  \;.
\label{2.1.8}
\end{eqnarray}
Considering that the overall phase of $M^{}_\nu$ (which is just the phase of $a a^\prime$) is of no physical meaning, we will take $a$ and $a^\prime$ to be real without affecting the physical results.
Such an $M^{}_\nu$ can be diagonalized by a TM1-type unitary matrix $U^{}_{\rm TM1}$ as follows
\begin{eqnarray}
U^{\dagger}_{\rm TM1} M^{}_\nu U^{*}_{\rm TM1} = D^{}_\nu = {\rm diag}(0, m^{}_2, m^{}_3) \;,
\label{2.1.9}
\end{eqnarray}
with
\begin{eqnarray}
U^{}_{\rm TM1} = \displaystyle \frac{1}{\sqrt 6} \left( \begin{array}{ccc}
2 & \sqrt{2} & 0 \cr
1 & - \sqrt{2}  & -\sqrt{3}  \cr
1 & - \sqrt{2}  & \sqrt{3} \cr
\end{array} \right) \left( \begin{array}{ccc}
1 & 0 & 0 \cr
0 & \cos \theta  & \sin \theta e^{-{\rm i} \varphi}  \cr
0 & - \sin \theta e^{{\rm i} \varphi}  & \cos \theta \cr
\end{array} \right)
\left( \begin{array}{ccc}
1 & 0 & 0 \cr
0 & e^{{\rm i} \beta}  & 0  \cr
0 & 0  & e^{{\rm i} \gamma} \cr
\end{array} \right) \;.
\label{2.1.10}
\end{eqnarray}
The parameters $\theta$ and $\varphi$ can be determined as
\begin{eqnarray}
&& \tan{2\theta}  =  \frac{ \sqrt{6} \left| 3(x + y) + 2 |x|^2 y + 2 x |y|^2 \right| } {4 \left| x y \right|^2 - 9  } \;, \nonumber \\
&&  \varphi =  \arg \left[ 3(x + y) + 2 |x|^2 y + 2 x |y|^2 \right] \;.
\label{2.1.11}
\end{eqnarray}
And the corresponding neutrino masses and the associated $\beta$ and $\gamma$ phases are given by
\begin{eqnarray}
&& m^{}_2 e^{2 {\rm i}\beta } = - \frac{a a^\prime}{M} \left[ 6 \cos^2 \theta + 4 x y \sin^2 \theta e^{-2{\rm i} \varphi} - \sqrt{6} (x + y) \sin 2\theta e^{-{\rm i} \varphi}  \right] \;, \nonumber \\
&& m^{}_3 e^{2 {\rm i}\gamma }  =  - \frac{a a^\prime}{M} \left[ 4 x y \cos^2 \theta + 6 \sin^2 \theta e^{2{\rm i} \varphi} + \sqrt{6} (x + y) \sin 2\theta e^{{\rm i} \varphi}  \right] \;.
\label{2.1.12}
\end{eqnarray}
Subsequently, the neutrino mixing angles and CP phases can be extracted from $U^{}_{\rm TM1}$ in Eq.~(\ref{2.1.10}) as
\begin{eqnarray}
& & s^{2}_{13} = \frac{1}{3} \sin^2 \theta \; , \hspace{1cm} s^{2}_{12} =  \frac{ 1}{3} -  \frac{2 s^{2}_{13}}{3 - 3s^{2}_{13}} \;, \hspace{1cm}
s^{2}_{23} = \frac{1}{2} + \frac{\sqrt{6} \sin 2\theta \cos \varphi}{6 - 2 \sin^2 \theta}  \;, \nonumber \\
&& \tan{2\theta^{}_{23}} \cos \delta = - \frac{1-5 s^2_{13}}{2 \sqrt{2} s^{}_{13} \sqrt{1- 3 s^2_{13}} } \;, \hspace{1cm}  \sigma =  \varphi - \delta + \beta - \gamma \;,
\label{2.1.13}
\end{eqnarray}
(note that the Majorana CP phase $\rho$ becomes unphysical due to the vanishing of $m^{}_1$).
With the help of Eq.~(\ref{2.1.13}), by inputting the $3\sigma$ ranges of $s^2_{13}$ and $s^2_{23}$, $\theta$ and $|\varphi|$ are respectively constrained into the ranges 0.25---0.27 and 0.33$\pi$---0.65$\pi$. And $s^2_{12}$ and $|\delta|$ are respectively predicted to lie in the ranges 0.347---0.350 and 0.34$\pi$---0.64$\pi$, which are in good agreement with the neutrino oscillation data.

With the help of Eqs.~(\ref{2.1.11}-\ref{2.1.13}) and the experimental results for $s^2_{13}$, $s^2_{23}$, $\Delta m^2_{21}$ and $\Delta m^2_{31}$, the allowed values of $a a^\prime/M$, $|x|$, $|y|$, $\arg{(x)}$ and $\arg{(y)}$ can be calculated. The results are shown in Figure~1. Since $M^{}_\nu$ in Eq.~(\ref{2.1.8}) keeps invariant with respect to the interchange $x\leftrightarrow y$, here we have only shown the results for $|x|>|y|$ while the results for $|x|<|y|$ can be simply obtained with the help of such an interchange.
The results show that only for $0.8 \lesssim a a^\prime/M \lesssim 1.9$ meV can the model considered be consistent with the experimental results. As can be seen from Eq.~(\ref{2.1.8}), this means that the effective Majorana neutrino mass $|(M^{}_\nu)^{}_{ee}|$ (i.e., the $ee$ element of $M^{}_\nu$) that controls the rates of neutrinoless double beta decays is constrained into the range 1.6---3.8 meV. Furthermore, $|x|$ lies between 2---10 while $|y|$ is around 2. And $\arg(x)$ is mainly in the range $-\pi$---0 while $\arg(y)$ is within the range 0---$\pi$. It is interesting to note that $\arg(x)$ and $\arg(y)$ have chance to take the special value $-\pi/2$ (corresponding to a purely imaginary $x$ or $y$) while $\arg(x)$ also has chance to take the special values 0 and $\pi$ (corresponding to a real $x$).

\begin{figure*}
\centering
\includegraphics[width=6.5in]{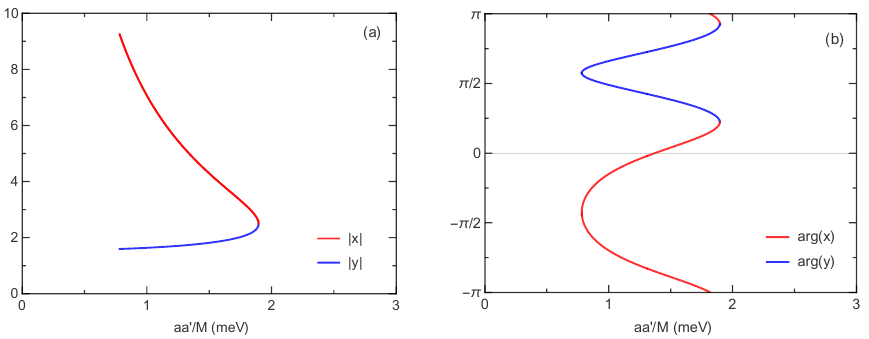}
\caption{ For the model given in section~2.1, the allowed values of $|x|$, $|y|$, $\arg{(x)}$ and $\arg{(y)}$ as functions of $a a^\prime/M$.  }
\end{figure*}

\subsection{Consequence for leptogenesis}

Before we study the consequence of the specific model given in last subsection [with $M^{}_{\rm R}$ in Eq.~(\ref{9}) and $M^{}_{\rm D}$ in Eq.~(\ref{2.1.5})] for leptogenesis, we first elucidate our common strategy for the leptogenesis calculations in the minimal seesaw model with $M^{}_{\rm R}$ in Eq.~(\ref{9}), which will also apply for the specific models studied in section~3.
First of all, as is usual in the literature, we go into the mass basis of right-handed neutrinos via the following unitary transformation:
\begin{eqnarray}
U^T_{\rm R} M^{}_{\rm R} U^{}_{\rm R} = {\rm diag}(M, M)  \hspace{0.5cm} {\rm with} \hspace{0.5cm}
U^{}_{\rm R} = \frac{1}{\sqrt 2} \left( \begin{array}{cc} 1 & 1 \cr - 1  & 1 \end{array} \right) P \; ,
\label{2.2.1}
\end{eqnarray}
where $P = {\rm diag} ( {\rm i}, 1 )$ serves to make the right-handed neutrino masses positive.
In the meantime, $M^{}_{\rm D}$ becomes
\begin{eqnarray}
M^{\prime}_{\rm D} = M^{}_{\rm D} U^{}_{\rm R} \;.
\label{2.2.2}
\end{eqnarray}
It should be noted that at the present stage the two right-handed neutrinos are degenerate in their masses, which prohibit leptogenesis to work. In order to have a successful leptogenesis, the degeneracy between the two right-handed neutrino masses needs to be broken. But in order to take advantage of the resonant leptogenesis scenario which can help us accommodate successful leptogenesis in the low-scale seesaw models, the splitting between the two right-handed neutrino masses should be tiny. In this paper, we consider two possible ways of generating the tiny splitting between the two right-handed neutrino masses: one way is to modify $M^{}_{\rm R}$ to a form as
\begin{eqnarray}
M^{}_{\rm R}= \displaystyle \left( \begin{array}{cc}
\mu & M \cr
M & \mu
\end{array} \right)  \;,
\label{2.2.3}
\end{eqnarray}
with $\mu \ll M$. Such a modification of $M^{}_{\rm R}$ has the merit that it only introduces one additional parameter and will not alter the unitary matrix $U^{}_{\rm R}$ in Eq.~(\ref{2.2.1}) for the purpose of its diagonalization. For such an $M^{}_{\rm R}$, the two right-handed neutrino masses are changed to $M-\mu$ and $M+\mu$ respectively (yielding $\Delta M \equiv M^{}_2 - M^{}_1 = 2\mu$). Another way is to consider the renormalization-group corrections for the right-handed neutrino masses \cite{rgeL}: if the energy scale $\Lambda^{}_{}$ of the flavor-symmetry physics that shapes the special textures of neutrino mass matrices is much higher than the right-handed neutrino mass scale $M$ where leptogenesis takes place, then the renormalization group evolution effects (between the scales of $\Lambda^{}_{}$ and $M$) may induce a considerable splitting between the two right-handed neutrino masses \cite{ynu}:
\begin{eqnarray}
\Delta M  \simeq \displaystyle \frac{M}{8\pi^2} \left[ (Y^{\prime\dagger}_\nu Y^{\prime}_\nu)^{}_{11} - (Y^{\prime\dagger}_\nu Y^{\prime}_\nu)^{}_{22} \right] \ln\left( \frac{\Lambda^{}_{}}{M} \right) \;,
\label{2.2.4}
\end{eqnarray}
with $Y^{\prime}_\nu = M^{\prime}_{\rm D}/v$.

As is well known, depending on the temperature ranges where leptogenesis takes place, different lepton flavors may become relevant \cite{flavor}. In the TeV-scale seesaw models considered in the present paper, all the $y^{}_\alpha$-related interactions (with $y^{}_\alpha$ being the charged-lepton Yukawa couplings) have entered thermal equilibrium, so all the three lepton flavors are distinguishable and should be treated separately. Furthermore, due to the quasi-degeneracy of the two right-handed neutrinos, their contributions to the final baryon asymmetry will be of equal importance. For these two reasons, the final baryon asymmetry can be calculated according to \cite{resonant}
\begin{eqnarray}
Y^{}_{\rm B} = c r \sum^{}_{\alpha} \varepsilon^{}_{\alpha}  \kappa ( \widetilde m^{}_\alpha) = c r \left[ \varepsilon^{}_{e}  \kappa ( \widetilde m^{}_e ) +
\varepsilon^{}_{\mu}  \kappa ( \widetilde m^{}_\mu ) + \varepsilon^{}_{\tau}  \kappa ( \widetilde m^{}_\tau )  \right] \;,
\label{2.2.5}
\end{eqnarray}
where $c = - 28/79$ describes the conversion efficiency from the lepton-antilepton asymmetry to the baryon-antibaryon asymmetry via the sphaleron processes, and $r \simeq 4 \times 10^{-3}$ measures the ratio of the number density of right-handed neutrinos to the entropy density. Then, the efficiency factor $\kappa \left( \widetilde m^{}_\alpha\right) <1$ takes account of the washout effects due to the inverse decay and various lepton-number-violating scattering processes, whose value relies on the following washout mass parameter and can be calculated by numerically solving relevant Boltzmann equations \cite{Lreview}
\begin{eqnarray}
\widetilde m^{}_\alpha = \widetilde m^{}_{1 \alpha} + \widetilde m^{}_{2 \alpha} \hspace{0.5cm} {\rm with} \hspace{0.5cm} \widetilde m^{}_{I \alpha} = \frac{|(M^{\prime}_{\rm D})^{}_{\alpha I}|^2}{M^{}_I} \;.
\label{2.2.6}
\end{eqnarray}
Finally, $\varepsilon^{}_{\alpha}$ is the sum of the $\alpha$-flavor CP asymmetries $\varepsilon^{}_{I\alpha}$ (i.e., $\varepsilon^{}_{\alpha} = \varepsilon^{}_{1\alpha} + \varepsilon^{}_{2\alpha}$), which measure the asymmetries between the decay rates of $N^{}_I \to L^{}_\alpha + H$ and their CP-conjugate processes $N^{}_I \to \overline{L}^{}_\alpha + \overline{H}$. In the resonant leptogenesis scenario, $\varepsilon^{}_{I\alpha}$ are given by \cite{resonant}
\begin{eqnarray}
\varepsilon^{}_{I\alpha} = \frac{{\rm Im}\left\{ (M^{\prime *}_{\rm D})^{}_{\alpha I} (M^{\prime}_{\rm D})^{}_{\alpha J}
\left[ M^{}_J (M^{\prime \dagger}_{\rm D} M^{\prime}_{\rm D})^{}_{IJ} + M^{}_I (M^{\prime \dagger}_{\rm D} M^{\prime}_{\rm D})^{}_{JI} \right] \right\} }{8\pi  v^2 (M^{\prime \dagger}_{\rm D} M^{\prime}_{\rm D})^{}_{II}} \cdot \frac{M^{}_I \Delta M^2_{IJ}}{(\Delta M^2_{IJ})^2 + M^2_I \Gamma^2_J} \;,
\label{2.2.7}
\end{eqnarray}
with $\Delta M^2_{IJ} \equiv M^2_I - M^2_J$ and $\Gamma^{}_J= (M^{\prime \dagger}_{\rm D} M^{\prime}_{\rm D})^{}_{JJ} M^{}_J/(8\pi v^2)$ being the decay rate of $N^{}_J$ (for $J \neq I$).

Now we are ready to study the consequence of the specific model given in last subsection for leptogenesis. For $M^{}_{\rm D}$ in Eq.~(\ref{2.1.5}), $M^{\prime}_{\rm D}$ is obtained as
\begin{eqnarray}
M^{\prime}_{\rm D} = M^{}_{\rm D} U^{}_{\rm R} = \displaystyle \frac{1}{\sqrt 2} \left( \begin{array}{cc}
{\rm i} (a - a^\prime) & a + a^\prime \cr
{\rm i} [- a (1+x) + a^\prime (1+y) ] & - a (1+x) - a^\prime (1+y) \cr
{\rm i} [- a (1-x) + a^\prime (1-y) ] & - a (1-x) - a^\prime (1-y)
\end{array} \right) \;.
\label{2.2.8}
\end{eqnarray}
For this $M^{\prime}_{\rm D}$, the washout mass parameters $\widetilde m^{}_{\alpha}$ and CP asymmetries
$\varepsilon^{}_{\alpha}$ are explicitly given by
\begin{eqnarray}
&& \widetilde m^{}_{e} = \frac{|a|^2 + |a^\prime|^2}{M} \;, \hspace{1cm} \varepsilon^{}_{e} =  (|a|^2 -|a^\prime|^2) \Theta \;, \nonumber \\
&& \widetilde m^{}_{\mu} =  \frac{|a(1+x)|^2 + |a^\prime (1+y)|^2}{M} \; , \hspace{1cm} \varepsilon^{}_{\mu} = (|a(1+x)|^2 - |a^\prime (1+y)|^2) \Theta  \;, \nonumber \\
&& \widetilde m^{}_{\tau} =  \frac{|a(1-x)|^2 + |a^\prime (1-y)|^2}{M} \;, \hspace{1cm}  \varepsilon^{}_{\tau} =  (|a(1-x)|^2 + |a^\prime (1-y)|^2) \Theta \;,
\label{2.2.10}
\end{eqnarray}
where $\Theta$ stands for
\begin{eqnarray}
\Theta = - {\rm Im}[a a^{\prime *} (3+2 x y^*)] \frac{M \Delta M}{4\pi v^2} \left\{  \frac{1}{ (M^{\prime \dagger}_{\rm D} M^{\prime}_{\rm D})^{}_{11} [4 (\Delta M)^2 + \Gamma^2_2]} + \frac{1}{ (M^{\prime \dagger}_{\rm D} M^{\prime}_{\rm D})^{}_{22} [4 (\Delta M)^2 + \Gamma^2_1]} \right\}
\label{2.2.11}
\end{eqnarray}
with
\begin{eqnarray}
(M^{\prime \dagger}_{\rm D} M^{\prime}_{\rm D})^{}_{11} = \frac{1}{2} \left\{ |a|^2 (3+2 |x|^2) + |a^\prime|^2 (3+2 |y|^2) - 2 {\rm Re}[ a a^{\prime *} (3+ 2x y^*)] \right\} \;, \nonumber \\
(M^{\prime \dagger}_{\rm D} M^{\prime}_{\rm D})^{}_{22} = \frac{1}{2} \left\{ |a|^2 (3+2 |x|^2) + |a^\prime|^2 (3+2 |y|^2) + 2 {\rm Re}[ a a^{\prime *} (3+2 x y^*)] \right\} \;.
\label{2.2.12}
\end{eqnarray}

We first consider the possibility that the splitting between the two right-handed neutrino masses is realized by modifying $M^{}_{\rm R}$ into the form as shown in Eq.~(\ref{2.2.3}). Figure~2(a) has shown the required values of $\mu$ (in order for leptogenesis to work successfully) as functions of $a a^\prime/M$.
In obtaining these results (and in the following numerical calculations) we have taken $M=1$ TeV as a benchmark value. In fact, the dependence of the final results on the values of $M$ is very weak, which is a generic feature of the resonant leptogenesis scenario \cite{resonant}. We have also taken some benchmark values for $a^\prime/a$. This is because although $M^{}_\nu$ is only dependent on the product of $a$ and $a^\prime$ [see Eq.~(\ref{2.1.8})], the leptogenesis results do depend on their individual values. The results show that, in order to achieve a successful leptogenesis, $\mu$ should lie in the range $\sim 10^{-4}$---$\sim 10^3$ eV, and $a^\prime/a$ should be a small quantity in the range $\sim 0.001$---$\sim 0.1$. It is interesting to note that a small $a^\prime/a$ is consistent with the linear seesaw model where an approximate lepton number conservation is invoked \cite{LNC}: the right-handed neutrinos $N^{}_1$ and $N^{}_2$ are endowed with the lepton numbers $+1$ and $-1$, respectively. When the lepton number is conserved exactly, $M^{}_{\rm R}$ is constrained into the form shown in Eq.~(\ref{9}), while the second column of $M^{}_{\rm D}$ (i.e., $a^\prime$) is simply vanishing. In this way the relative smallness of $a^\prime$ compared to $a$ can be naturally interpreted as a result of the lepton number conservation being broken but only to a small degree.

\begin{figure*}
\centering
\includegraphics[width=6.5in]{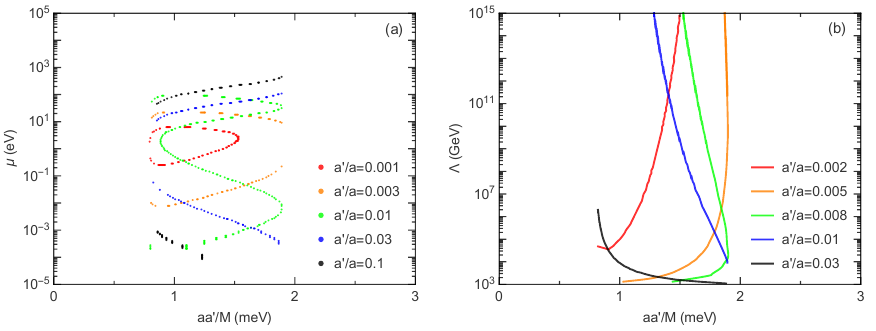}
\caption{ (a) For the model given in section~2.1, the values of $\mu$ that allow for a successful leptogenesis as functions of $a a^\prime/M$, in the scenario that the splitting between the two right-handed neutrino masses is realized by modifying $M^{}_{\rm R}$ into the form as shown in Eq.~(\ref{2.2.3}).  (b) The values of $\Lambda^{}_{}$ that allow for a successful leptogenesis, in the scenario that the splitting between the two right-handed neutrino masses is generated from the renormalization-group corrections.}
\end{figure*}

We then consider the possibility that the splitting between the two right-handed neutrino masses is generated from the renormalization-group corrections as shown in Eq.~(\ref{2.2.4}). Similar to Figure~2(a), Figure~2(b) has shown the required values of $\Lambda$ (in order for leptogenesis to work successfully) as functions of $a a^\prime/M$. The results show that successful leptogenesis can be achieved for $\Lambda$ ranging from TeV to the grand unification scale, and $a^\prime/a$ in the range $\sim 0.001$---$\sim 0.03$.

\section{Realization and consequence of the $\mu$-$\tau$ reflection symmetry}

In this section, we study the realization and consequence of the $\mu$-$\tau$ reflection symmetry in the minimal seesaw model with $M^{}_{\rm R}$ in Eq.~(\ref{9}).

\subsection{Realization of the $\mu$-$\tau$ reflection symmetry}

We first consider the realization of the $\mu$-$\tau$ reflection symmetry in the minimal seesaw model with $M^{}_{\rm R}$ in Eq.~(\ref{9}). In order for $M^{}_\nu$ in Eq.~(\ref{2.1.2}) to respect the $\mu$-$\tau$ reflection symmetry as in Eq.~(\ref{5}), one has the following two options:
\begin{eqnarray}
&& {\rm Case \ I:} \hspace{1cm} a^\prime = a^* \;, \hspace{1cm} b^\prime = c^*  \;, \hspace{1cm}  c^\prime = b^{*}  \;; \nonumber \\
&& {\rm Case \ II:} \hspace{1cm} {\rm Im}(a) = {\rm Im}(a^\prime) =0 \;, \hspace{1cm} c = b^*  \;, \hspace{1cm}  c^\prime = b^{\prime *}  \;.
\label{3.1.1}
\end{eqnarray}
Therefore, in the minimal seesaw model with $M^{}_{\rm R}$ in Eq.~(\ref{9}), the form of $M^{}_{\rm D}$ that naturally realizes the $\mu$-$\tau$ reflection symmetry can be expressed as
\begin{eqnarray}
{\rm Case \ I:} \hspace{1cm} M^{\rm I}_{\rm D}= \displaystyle \left( \begin{array}{cc}
a & a^* \cr
b & c^* \cr
c & b^*
\end{array} \right)\;; \hspace{1.5cm}
{\rm Case \ II:} \hspace{1cm} M^{\rm II}_{\rm D}= \displaystyle \left( \begin{array}{cc}
a & a^\prime \cr
b & b^\prime \cr
b^* & b^{\prime *}
\end{array} \right)  \;,
\label{3.1.2}
\end{eqnarray}
with $a$ and $a^\prime$ being real in Case II and all the other parameters being complex.

Now, for $M^{}_{\rm R}$ in Eq.~(\ref{9}) and $M^{}_{\rm D}$ in Eq.~(\ref{3.1.2}), the seesaw formula gives an $M^{}_\nu$ as
\begin{eqnarray}
&& {\rm Case \ I:} \hspace{1cm} M^{\rm I}_\nu =  \displaystyle - \frac{1}{M} \left( \begin{array}{ccc}
2 |a|^2 & a c^* + a^* b & a b^* + a^* c \cr
a c^* + a^* b & 2 b c^*  & |b|^2+|c|^2  \cr
a b^* + a^* c & |b|^2+|c|^2  & 2 b^* c \cr
\end{array} \right)  \;; \nonumber \\
&& {\rm Case \ II:} \hspace{1cm} M^{\rm II}_\nu =  \displaystyle - \frac{1}{M} \left( \begin{array}{ccc}
2 a a^\prime & a b^\prime + b a^\prime & a b^{\prime *} + b^* a^\prime \cr
a b^\prime + b a^\prime & 2 b b^\prime  & b b^{\prime *} + b^* b^\prime  \cr
a b^{\prime *} + b^* a^\prime & b b^{\prime *} + b^* b^\prime  & 2 b^* b^{\prime *} \cr
\end{array} \right)  \;.
\label{3.1.3}
\end{eqnarray}
Both $M^{\rm I}_\nu$ and $M^{\rm II}_\nu$ can be diagonalized by a unitary matrix as follows
\begin{eqnarray}
U^{}_{\mu\tau}  = \frac{1}{\sqrt{2}}
\left( \begin{matrix}
1 &  & \cr
& e^{{\rm i} \phi}  & \cr
&  & -e^{-{\rm i} \phi}
\end{matrix} \right)
 \left( \begin{matrix}
\sqrt{2} c^{}_{12} c^{}_{13} & \sqrt{2} s^{}_{12} c^{}_{13} & - \sqrt{2} {\rm i} \eta^{}_\delta s^{}_{13}  \cr
-s^{}_{12}- {\rm i} \eta^{}_\delta c^{}_{12} s^{}_{13}
& c^{}_{12} - {\rm i} \eta^{}_\delta s^{}_{12} s^{}_{13}   & c^{}_{13} \cr
s^{}_{12} - {\rm i} \eta^{}_\delta c^{}_{12} s^{}_{13}
& -c^{}_{12} - {\rm i} \eta^{}_\delta s^{}_{12} s^{}_{13}  & c^{}_{13}
\end{matrix} \right) \left( \begin{matrix}
e^{{\rm i}\alpha } &  & \cr
& e^{{\rm i}\beta }  & \cr
&  & e^{ {\rm i}\gamma }
\end{matrix} \right) \;,
\label{3.1.4}
\end{eqnarray}
with $\eta^{}_{\delta} = \pm 1$ for $\delta =\pm \pi/2$.

In Case I, the parameters $\theta^{}_{12}$, $\theta^{}_{13}$ and $\phi$ of $U^{}_{\mu\tau}$ are determined by
\begin{eqnarray}
&& \tan \theta^{}_{13} = - \frac{\sqrt{2} \eta^{}_\delta  {\rm Im}[b c^* e^{-2{\rm i}\phi} ] }{ {\rm Re}[(a c^* + a^* b )e^{-{\rm i} \phi} ] } \;, \hspace{1cm} \tan 2 \theta^{}_{13} = - \frac{2\sqrt{2} \eta^{}_\delta {\rm Im}[ (a c^* + a^* b )e^{-{\rm i} \phi} ] }{ 2|a|^2 + 2 {\rm Re}( b c^* e^{-2{\rm i} \phi }) - |b|^2 -|c|^2 }  \;, \nonumber \\
&&  \tan 2 \theta^{}_{12}   =  \frac{ 2 \Delta^{}_2 }{ \Delta^{}_3- \Delta^{}_1 } \;,
\label{3.1.5}
\end{eqnarray}
with
\begin{eqnarray}
&& \Delta^{}_1 =  -2 c^2_{13} |a|^2  + s^2_{13} [ 2 {\rm Re}( b c^* e^{-2{\rm i} \phi }) - |b|^2 - |c|^2 ] + 2 \sqrt{2} \eta^{}_\delta c^{}_{13} s^{}_{13} {\rm Im}[ (a c^* + a^* b )e^{-{\rm i} \phi} ] \;, \nonumber  \\
&& \Delta^{}_2 = - \sqrt{2} c^{}_{13} {\rm Re}[(a c^* + a^* b) e^{-{\rm i}\phi} ] + 2 \eta^{}_{\delta} s^{}_{13}{\rm Im}(b c^* e^{-2{\rm i} \phi})  \;,  \nonumber  \\
&& \Delta^{}_3 = - 2 {\rm Re}(b c^* e^{-2{\rm i} \phi}) - |b|^2 -|c|^2 \;.
\label{3.1.6}
\end{eqnarray}
And the corresponding neutrino masses are given by
\begin{eqnarray}
&&  m^{}_1 e^{2 {\rm i}\alpha }  = \frac{1}{M} (c^2_{12} \Delta^{}_1 + s^2_{12} \Delta^{}_3 - 2 c^{}_{12} s^{}_{12} \Delta^{}_2)  \;, \hspace{1cm} m^{}_2 e^{2 {\rm i}\beta } = \frac{1}{M} (s^2_{12} \Delta^{}_1 + c^2_{12} \Delta^{}_3 + 2 c^{}_{12} s^{}_{12} \Delta^{}_2 ) \;, \nonumber  \\
&& m^{}_3 e^{2 {\rm i}\gamma } =  \frac{1}{M} \left\{ c^2_{13} [ |b|^2+|c|^2 - 2 {\rm Re}(b c^* e^{-2{\rm i} \phi}) ] + 2 s^2_{13} |a|^2 + 2 \sqrt{2} \eta^{}_\delta c^{}_{13} s^{}_{13} {\rm Im}[ (a c^* + a^* b )e^{-{\rm i} \phi} ] \right\} \;.
\label{3.1.7}
\end{eqnarray}
Note that one of $m^{}_1$ and $m^{}_3$ will be vanishing, depending on the parameter values.
Obviously, since $\Delta^{}_1$, $\Delta^{}_2$ and $\Delta^{}_3$ are real, one has $\alpha, \beta, \gamma= 0$ or $\pi/2$. From these results $\rho$ and $\sigma$ can be subsequently obtained as $\rho = \alpha - \gamma$ and $\sigma = \beta -\gamma$.

With the help of Eqs.~(\ref{3.1.5}-\ref{3.1.7}) and the experimental results for $\theta^{}_{12}$, $\theta^{}_{13}$, $\Delta m^2_{21}$ and $\Delta m^2_{31}$, the allowed values of the model parameters in $M^{\rm I}_{\rm D}$ can be calculated. Note that $M^{\rm I}_\nu$ only depends on the combinations $\arg(a^* b) = \arg{(b)} -\arg{(a)} $ and $\arg(a^* c) =\arg{(c)} -\arg{(a)}$ among $\arg{(a)}$, $\arg{(b)}$ and $\arg{(c)}$, so only $|a|$, $|b|$, $|c|$, $\arg(a^* b) $ and $\arg(a^* c)$ can be determined. Furthermore, since $M^{\rm I}_\nu$ keeps invariant with respect to the interchange between $b\leftrightarrow c^*$, in the following we will only show the results for $|b|>|c|$ while the results for $|b|<|c|$ can be simply obtained with the help of such an interchange.
In the NO case (with $m^{}_1=0$, where the Majorana CP phase $\rho$ becomes unphysical), the results are given by
\begin{eqnarray}
&& \sigma=0: \hspace{1cm} |a| =  2.7 \times 10^4 \ {\rm eV} \;, \hspace{0.5cm} |b|= 2.7 \times 10^5 \ {\rm eV} \;,  \hspace{0.5cm} |c|= 5.1 \times 10^4 \ {\rm eV} \;, \nonumber \\
&& \hspace{2.3cm} \arg(a^* b) = 0.40 \pi \;, \hspace{0.5cm} \arg(a^* c) = 0.39\pi \;; \nonumber \\
&& \sigma=\frac{\pi}{2}: \hspace{1cm} |a| = 4.3 \times 10^4 \ {\rm eV}  \;, \hspace{0.5cm} |b|= 1.5 \times 10^5 \ {\rm eV}  \;,  \hspace{0.5cm} |c|= 7.2 \times 10^4 \ {\rm eV}  \;,\nonumber \\
&& \hspace{2.3cm} \arg(a^* b) = -0.18 \pi \;, \hspace{0.5cm} \arg(a^* c) = 0.83 \pi \;.
\label{3.1.8}
\end{eqnarray}
In obtaining these results, we have taken $M=1$ TeV as a benchmark value. For other values of $M$,
the results of $|a|$, $|b|$ and $|c|$ can be obtained with the help of a simple rescaling law, while the results of $\arg{(a^* b)}$ and $\arg{(a^*c)}$ keep invariant. Furthermore, in the numerical calculations, we have fixed $\delta = -\pi/2$ out of $\pm \pi/2$ which is more experimentally favored. From these results, it is direct to obtain the effective Majorana neutrino mass $|(M^{}_\nu)^{}_{ee}|$ that controls the rates of neutrinoless double beta decays to be 1.4 or 3.7 meV in the case of $\sigma=0$ or $\pi/2$.
In the IO case (with $m^{}_3=0$, where only the difference $\rho-\sigma$ between the two Majorana CP phases is of physical meaning), the results are given by
\begin{eqnarray}
&& \rho-\sigma=0: \hspace{1cm} |a| = 1.6 \times 10^5 \ {\rm eV} \;, \hspace{0.5cm} |b|= 1.3 \times 10^5 \ {\rm eV}  \;,  \hspace{0.5cm} |c|= 9.5 \times 10^4 \ {\rm eV}  \;, \nonumber \\
&& \hspace{3cm} \arg(a^* b) = 0.50 \pi \;, \hspace{0.5cm} \arg(a^* c) = 0.50 \pi \;; \nonumber \\
&& \rho- \sigma=\frac{\pi}{2}: \hspace{1cm} |a| = 9.7 \times 10^4 \ {\rm eV} \;, \hspace{0.5cm} |b|= 3.4 \times 10^5 \ {\rm eV} \;,  \hspace{0.5cm} |c|= 1.8 \times 10^4 \ {\rm eV} \;, \nonumber \\
&& \hspace{3cm} \arg(a^* b) = 0.99 \pi \;, \hspace{0.5cm} \arg(a^* c) = -0.20 \pi \;.
\label{3.1.9}
\end{eqnarray}
And $|(M^{}_\nu)^{}_{ee}|$ is obtained to be 51.2 or 18.8 meV in the case of $\rho-\sigma=0$ or $\pi/2$.

In Case II, the parameters $\theta^{}_{13}$ and $\phi$ of $U^{}_{\mu\tau}$ are determined by
\begin{eqnarray}
&& \tan \theta^{}_{13} = - \frac{\sqrt{2} \eta^{}_\delta  {\rm Im}[b b^\prime e^{-2{\rm i}\phi} ] }{ {\rm Re}[(a b^\prime + b a^\prime )e^{-{\rm i} \phi} ] } \;, \hspace{1cm} \tan 2 \theta^{}_{13} = - \frac{\sqrt{2} \eta^{}_\delta {\rm Im}[ (a b^\prime + b a^\prime )e^{-{\rm i} \phi} ] }{ a a^\prime + {\rm Re}( b b^\prime e^{-2{\rm i} \phi }) - {\rm Re}( b b^{\prime *}) } \;,
\label{3.1.10}
\end{eqnarray}
while $\theta^{}_{12}$ is same as in Eq.~(\ref{3.1.5}) but with
\begin{eqnarray}
&& \Delta^{}_1 =  -2 c^2_{13} a a^\prime  + 2 s^2_{13} [ {\rm Re}( b b^\prime e^{-2{\rm i} \phi }) - {\rm Re}( b b^{\prime *}) ] + 2 \sqrt{2} \eta^{}_\delta c^{}_{13} s^{}_{13} {\rm Im}[ (a b^\prime + b a^\prime )e^{-{\rm i} \phi} ] \;, \nonumber  \\
&& \Delta^{}_2 = - \sqrt{2} c^{}_{13} {\rm Re}[(a b^\prime + b a^\prime) e^{-{\rm i}\phi} ] + 2 \eta^{}_{\delta} s^{}_{13}{\rm Im}(b b^\prime e^{-2{\rm i} \phi})  \;,  \nonumber  \\
&& \Delta^{}_3 = - 2 {\rm Re}(b b^\prime e^{-2{\rm i} \phi}) - 2 {\rm Re}(b b^{\prime *}) \;.
\label{3.1.11}
\end{eqnarray}
And the expressions for $m^{}_1 e^{2 {\rm i}\alpha }$ and $m^{}_3 e^{2 {\rm i}\gamma }$ are same as in Eq.~(\ref{3.1.7}) while the expression for $m^{}_3 e^{2 {\rm i}\gamma }$ becomes
\begin{eqnarray}
&& m^{}_3 e^{2 {\rm i}\gamma } =  \frac{1}{M} \left\{ 2 c^2_{13} [ {\rm Re}(b b^{\prime *}) - {\rm Re}(b b^\prime e^{-2{\rm i} \phi}) ] + 2 s^2_{13} a a^\prime + 2 \sqrt{2} \eta^{}_\delta c^{}_{13} s^{}_{13} {\rm Im}[ (a b^\prime + b a^\prime )e^{-{\rm i} \phi} ] \right\} \;.
\label{3.1.12}
\end{eqnarray}

\begin{figure*}
\centering
\includegraphics[width=6.5in]{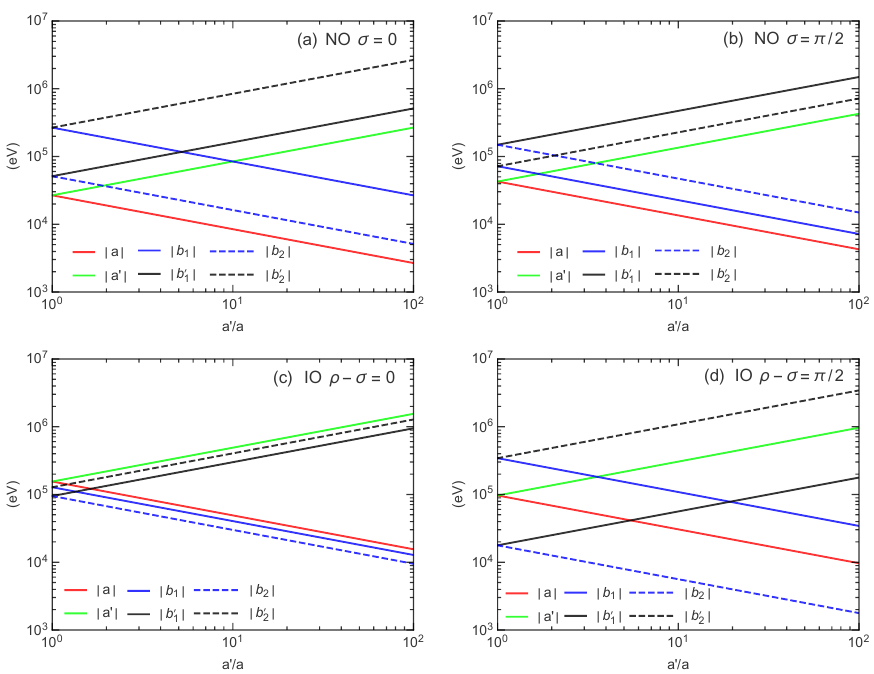}
\caption{ For the model with $M^{\rm II}_{\rm D}$ given in section~3.1, the allowed values of $|a|$, $|b|$, $|a^\prime|$ and $|b^\prime|$  as functions of $a^\prime/a$. (a) In the NO case with $\sigma=0$. (b) In the NO case with $\sigma=\pi/2$. (c) In the IO case with $\rho-\sigma=0$. (d) In the IO case with $\rho-\sigma=\pi/2$.   }
\end{figure*}

With the help of Eqs.~(\ref{3.1.10}-\ref{3.1.12}) and the experimental results for $\theta^{}_{12}$, $\theta^{}_{13}$, $\Delta m^2_{21}$ and $\Delta m^2_{31}$, the allowed values of the model parameters in $M^{\rm II}_{\rm D}$ can be calculated. Figure~3 has shown the allowed values of $|a|$, $|b|$, $|a^\prime|$ and $|b^\prime|$ as functions of $a^\prime/a$. Note that since $M^{\rm II}_\nu$ keeps invariant with respect to the joint interchanges $a \leftrightarrow a^\prime$ and $b \leftrightarrow b^\prime$, here we have only shown the results for $|a^\prime|>|a|$ while the results for $|a^\prime|<|a|$ can be simply obtained with the help of such interchanges. Furthermore, corresponding to each value combination of $|a|$ and $|a^\prime|$, there are two value combinations of $|b|$ and $|b^\prime|$. In order to distinguish these two value combinations, we have used $|b^{}_1|/|b^\prime_1|$ (in solid lines) and $|b^{}_2|/|b^\prime_2|$ (in dashed lines) to denote them.

\subsection{Consequence for leptogenesis}

Then, we study the consequence of the specific model given in last subsection [with $M^{}_{\rm R}$ in Eq.~(\ref{9}) and $M^{}_{\rm D}$ in Eq.~(\ref{3.1.2})] for leptogenesis. For $M^{}_{\rm D}$ in Eq.~(\ref{3.1.2}), $M^{\prime}_{\rm D}$ is obtained as
\begin{eqnarray}
&& {\rm Case \ I:} \hspace{1cm} M^{{\rm I}\prime}_{\rm D} = M^{\rm I}_{\rm D} U^{}_{\rm R} = \displaystyle \frac{1}{\sqrt 2} \left( \begin{array}{cc}
- 2 {\rm Im}(a) & 2 {\rm Re}(a) \cr
{\rm i} ( b - c^* ) & b + c^*  \cr
{\rm i} ( c - b^* ) & b^*+c
\end{array} \right) \;; \nonumber \\
&& {\rm Case \ II:} \hspace{1cm} M^{{\rm II}\prime}_{\rm D} = M^{\rm II}_{\rm D} U^{}_{\rm R} = \displaystyle \frac{1}{\sqrt 2} \left( \begin{array}{cc}
{\rm i} ( a - a^\prime) & a + a^\prime \cr
{\rm i} ( b - b^\prime ) & b + b^\prime  \cr
{\rm i} ( b^* - b^{\prime *} ) & b^* + b^{\prime *}
\end{array} \right)  \;.
\label{3.2.1}
\end{eqnarray}
In Case I, the washout mass parameters $\widetilde m^{}_{\alpha}$ and CP asymmetries $\varepsilon^{}_{\alpha}$ are explicitly given by
\begin{eqnarray}
&& \widetilde m^{}_{e} = \frac{2 |a|^2}{M} \;, \hspace{1cm} \widetilde m^{}_{\mu} =  \widetilde m^{}_{\tau} = \frac{|b|^2 + |c|^2}{M} \;, \hspace{1cm} \varepsilon^{}_{e} =0 \;, \nonumber \\
& & \varepsilon^{}_{\mu} = - \varepsilon^{}_{\tau} =
- (|b|^2 - |c|^2) [{\rm Re}(a) {\rm Im}(a) + {\rm Im}(bc)] \frac{M \Delta M}{2 \pi v^2} \nonumber \\ && \hspace{2.1cm}  \times \left\{  \frac{1}{ (M^{ {\rm I}\prime \dagger}_{\rm D} M^{{\rm I} \prime}_{\rm D})^{}_{11} [4 (\Delta M)^2 + \Gamma^2_2]} + \frac{1}{ (M^{{\rm I}\prime \dagger}_{\rm D} M^{{\rm I} \prime}_{\rm D})^{}_{22} [4 (\Delta M)^2 + \Gamma^2_1]} \right\} \;,
\label{3.2.2}
\end{eqnarray}
with
\begin{eqnarray}
&& (M^{ {\rm I} \prime \dagger}_{\rm D} M^{{\rm I} \prime}_{\rm D})^{}_{11} = 2 [{\rm Im}(a)]^2 + |b|^2 + |c|^2 - 2{\rm Re}(bc) \;, \nonumber \\
&& (M^{{\rm I} \prime \dagger}_{\rm D} M^{ {\rm I} \prime}_{\rm D})^{}_{22} = 2 [{\rm Re}(a)]^2 + |b|^2 + |c|^2 + 2{\rm Re}(bc) \;.
\label{3.2.3}
\end{eqnarray}
This means that the contributions of the $\mu$ and $\tau$ flavors to the baryon asymmetry exactly cancel out each other (due to $\varepsilon^{}_{\mu} = - \varepsilon^{}_{\tau}$ and $\widetilde m^{}_{\mu} =  \widetilde m^{}_{\tau}$), while the $e$ flavor has no contribution (due to $\varepsilon^{}_{e} =0$). Therefore, in the present case, in order for leptogenesis to work, one not only needs to generate a splitting between the two right-handed neutrino masses but also needs to break the exact cancellation between the contributions of the $\mu$ and $\tau$ flavors to the baryon asymmetry.
Note that in Eq.~(\ref{2.2.5}) we have used a flavor-universal conversion factor (i.e., $c=-28/79$) for the lepton-antilepton asymmetries in three lepton flavors to the baryon-antibaryon asymmetry via the sphaleron processes. However, as shown in Ref.~\cite{sphaleron}, the hierarchies in the charged-lepton Yukawa couplings lead to different conversion factors for different lepton flavors, although the differences among them are very tiny. This is just what we need to break the exact cancellation between the contributions of the $\mu$ and $\tau$ flavors to the baryon asymmetry: to be explicit, in Case I, after taking account of such an effect, the final baryon asymmetry is given by \cite{sphaleron}
\begin{eqnarray}
&& Y^{}_{\rm B} = r \left[ (c- 0.03 y^2_e) \varepsilon^{}_{e}  \kappa ( \widetilde m^{}_e ) +
(c- 0.03 y^2_\mu) \varepsilon^{}_{\mu}  \kappa ( \widetilde m^{}_\mu ) + (c- 0.03 y^2_\tau) \varepsilon^{}_{\tau}  \kappa ( \widetilde m^{}_\tau )  \right] \nonumber \\
&& \hspace{0.6cm} \simeq - 0.03 y^2_\tau r \varepsilon^{}_{\tau}  \kappa ( \widetilde m^{}_\tau ) \simeq  - 3 \times 10^{-6} r \varepsilon^{}_{\tau}  \kappa ( \widetilde m^{}_\tau )  \;.
\label{3.2.4}
\end{eqnarray}

Now, we first consider the possibility that the splitting between the two right-handed neutrino masses is realized by modifying $M^{}_{\rm R}$ into the form as shown in Eq.~(\ref{2.2.3}). Note that such a modification has the merit that it does not jeopardize the already-established $\mu$-$\tau$ reflection symmetry. Figure~4 has shown the maximally allowed values of $Y^{}_{\rm B}$ as functions of $\mu$, which are obtained by freely varying the values of $\arg(a)$ [which has not been constrained by the neutrino oscillation data as discussed above Eq.~(\ref{3.1.8})]. Unfortunately, the results show that in the present case the observed value of $Y^{}_{\rm B}$ cannot be successfully reproduced. In the NO case with $\sigma =0$, $Y^{}_{\rm B}$ can reach $3 \times 10^{-11}$ at most, smaller than its observed value by a factor of about 3. In the IO case, $Y^{}_{\rm B}$ is smaller than its observed value by about two orders of magnitude at least.

\begin{figure*}
\centering
\includegraphics[width=6.5in]{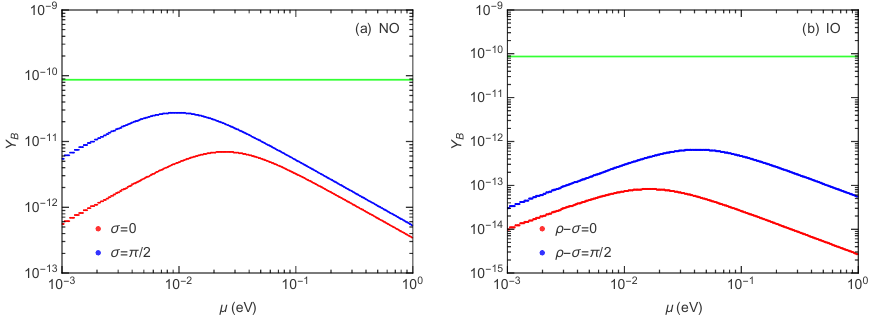}
\caption{ For the model with $M^{\rm I}_{\rm D}$ given in section~3.1, the maximally allowed values of $Y^{}_{\rm B}$ as functions of $\mu$, in the scenario that the splitting between the two right-handed neutrino masses is realized by modifying $M^{}_{\rm R}$ into the form as shown in Eq.~(\ref{2.2.3}).
(a) In the NO case with $\sigma =0$ (in the red color) and $\pi/2$ (in the blue color). (b) In the IO case with $\rho-\sigma=0$ (in the red color) and $\pi/2$ (in the blue color). The green horizontal line indicates the observed value of $Y^{}_{\rm B}$.  }
\end{figure*}

Then, we consider the possibility that the splitting between the two right-handed neutrino masses is generated from the renormalization-group corrections as shown in Eq.~(\ref{2.2.4}). But it should be noted that the renormalization group evolution effects also render the breaking of the $\mu$-$\tau$ reflection symmetry (and consequently the exact cancellation between the contributions of the $\mu$ and $\tau$ flavors to the baryon asymmetry), making another contribution to the final baryon asymmetry. This is because, due to the differences among the charged-lepton Yukawa couplings $y^{}_\alpha$, the Dirac neutrino mass matrix $M^{}_{\rm D}(M)$ at the right-handed neutrino mass scale $M$ will be corrected by the renormalization group evolution effects to the following form from its counterpart $M^{}_{\rm D}(\Lambda^{}_{})$ at the flavor-symmetry scale \cite{IRGE}
\begin{eqnarray}
M^{}_{\rm D} (M) \propto \left( \begin{array}{ccc}
1+\Delta^{}_{e} &   &  \cr
 & 1 +\Delta^{}_{\mu} &  \cr
 &  &  1+\Delta^{}_{\tau} \cr
\end{array} \right)
M^{}_{\rm D} (\Lambda^{}_{}) \;,
\label{3.2.5}
\end{eqnarray}
with
\begin{eqnarray}
\Delta^{}_{\alpha}   =   \frac{3}{32 \pi^2}\int^{\ln (\Lambda^{}_{}/M)}_{0} y^2_{\alpha} \ {\rm dt} \simeq \frac{3}{32 \pi^2} y^2_{\alpha} \ln \left(\frac{\Lambda^{}_{}}{M} \right) \;.
\label{3.2.6}
\end{eqnarray}
Before proceeding, we point out that, owing to $\Delta^{}_{e} \ll \Delta^{}_{\mu} \ll \Delta^{}_{\tau} \ll 1$, it is an excellent approximation for us to only keep $\Delta^{}_{\tau}$ in the following calculations. And one has $\Delta^{}_\tau \simeq 2 \times 10^{-6}$ in the SM, for $\Lambda^{}_{}/M =10$ as a benchmark value. Now that the $\mu$-$\tau$ reflection symmetry has broken, one obtains
\begin{eqnarray}
\varepsilon^{}_{\tau} \simeq - (1+ 2\Delta^{}_\tau) \varepsilon^{}_{\mu} \;, \hspace{1cm} \widetilde m^{}_\tau \simeq (1+ 2\Delta^{}_\tau) \widetilde m^{}_\mu \;.
\label{3.2.7}
\end{eqnarray}
Consequently, the final baryon asymmetry is given by
\begin{eqnarray}
&& Y^{}_{\rm B} =  c r \varepsilon^{}_{\mu} \left\{ \kappa ( \widetilde m^{}_\mu ) - (1+ 2\Delta^{}_\tau) \kappa \left[ (1+ 2\Delta^{}_\tau) \widetilde m^{}_\mu  \right]    \right\} \nonumber \\
&& \hspace{0.55cm} \simeq -c r \varepsilon^{}_{\mu} \left\{ 2\Delta^{}_\tau \kappa ( \widetilde m^{}_\mu ) + \kappa \left[ (1+ 2\Delta^{}_\tau) \widetilde m^{}_\mu  \right]  - \kappa ( \widetilde m^{}_\mu )  \right\} \;,
\label{3.2.8}
\end{eqnarray}
which does not suffer the exact cancellation between the contributions of the $\mu$ and $\tau$ flavors any more. Taking account of both the contributions from Eq.~(\ref{3.2.4}) and Eq.~(\ref{3.2.8}) to the final baryon asymmetry, Figure~5 has shown the maximally allowed values of $Y^{}_{\rm B}$ as functions of $\Lambda$. Unfortunately, the results show that in the present case the observed value of $Y^{}_{\rm B}$ cannot be successfully reproduced either. Similar to the results in the scenario that the splitting between the two right-handed neutrino masses is realized by modifying $M^{}_{\rm R}$ into the form as shown in Eq.~(\ref{2.2.3}), in the NO case with $\sigma =\pi/2$, $Y^{}_{\rm B}$ can reach $2 \times 10^{-11}$ at most, smaller than its observed value by a factor of about 4. In the IO case, $Y^{}_{\rm B}$ is smaller than its observed value by about three orders of magnitude or even worse.

\begin{figure*}
\centering
\includegraphics[width=6.5in]{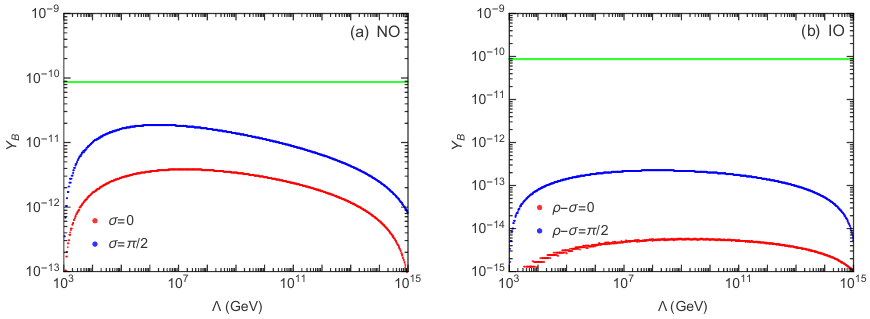}
\caption{ Same as Figure~4, except that here shown are the maximally allowed values of $Y^{}_{\rm B}$ as functions of $\Lambda$, in the scenario that the splitting between the two right-handed neutrino masses is generated from the renormalization-group corrections.  }
\end{figure*}

In Case II, the washout mass parameters $\widetilde m^{}_{\alpha}$ are explicitly given by
\begin{eqnarray}
&& \widetilde m^{}_{e} = \frac{a^2 + a^{\prime 2}}{M} \;, \hspace{1cm} \widetilde m^{}_{\mu} =  \widetilde m^{}_{\tau} = \frac{|b|^2 + |b^\prime|^2}{M} \;,
\label{3.2.9}
\end{eqnarray}
while the CP asymmetries $\varepsilon^{}_{\alpha}$ are simply vanishing. This means that, in the present case, leptogenesis cannot work through the mechanism described by Eq.~(\ref{3.2.4}) unless the $\mu$-$\tau$ reflection symmetry gets broken. Therefore, we will not consider the possibility that the splitting between the two right-handed neutrino masses is realized by modifying $M^{}_{\rm R}$ into the form as shown in Eq.~(\ref{2.2.3}), which keeps the $\mu$-$\tau$ reflection symmetry intact.
Now we consider the possibility that the splitting between the two right-handed neutrino masses is generated from the renormalization-group corrections as shown in Eq.~(\ref{2.2.4}), along with which the renormalization group evolution effects also lead to the breaking of the $\mu$-$\tau$ reflection symmetry as shown in Eq.~(\ref{3.2.5}). Thanks to the renormalization group evolution effects,
$\varepsilon^{}_{\alpha}$ now become non-vanishing as
\begin{eqnarray}
&& \varepsilon^{}_{e} \simeq \Delta^{}_{\tau} (a^2 - a^{\prime2})  {\rm Im}(bb^{\prime *}) \frac{M \Delta M}{2 \pi v^2} \left\{  \frac{1}{ (M^{ {\rm II} \prime \dagger}_{\rm D} M^{{\rm II} \prime}_{\rm D})^{}_{11} [4 (\Delta M)^2 + \Gamma^2_2]} + \frac{1}{ (M^{{\rm II} \prime \dagger}_{\rm D} M^{{\rm II} \prime}_{\rm D})^{}_{22} [4 (\Delta M)^2 + \Gamma^2_1]} \right\} \;, \nonumber \\
&& \varepsilon^{}_{\tau} \simeq \varepsilon^{}_{\mu} \simeq \frac{|b|^2 - |b^\prime|^2}{a^2 - a^{\prime2}} \varepsilon^{}_{e} \;,
\label{3.2.2}
\end{eqnarray}
with
\begin{eqnarray}
&& (M^{ {\rm II} \prime \dagger}_{\rm D} M^{ {\rm II} \prime}_{\rm D})^{}_{11} = \frac{1}{2} (a-a^\prime)^2 + |b|^2 + |b^\prime|^2 - 2{\rm Re}(bb^{\prime *}) \;, \nonumber \\
&& (M^{ {\rm II} \prime \dagger}_{\rm D} M^{{\rm II} \prime}_{\rm D})^{}_{22} = \frac{1}{2} (a+a^\prime)^2 + |b|^2 + |b^\prime|^2 + 2{\rm Re}(bb^{\prime *}) \;.
\label{3.2.3}
\end{eqnarray}
Similar to Figure~5, Figure~6 has shown the maximally allowed values of $Y^{}_{\rm B}$ as functions of $\Lambda$. We see that in the NO case with $\sigma =\pi/2$, $Y^{}_{\rm B}$ can reach $5 \times 10^{-11}$ at most, smaller than its observed value by a factor of about 2. In the IO case, $Y^{}_{\rm B}$ is smaller than its observed value by about three orders of magnitude.

\begin{figure*}
\centering
\includegraphics[width=6.5in]{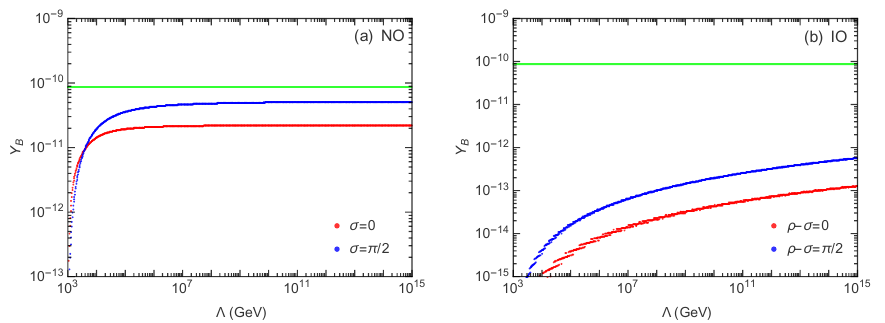}
\caption{ Same as Figure~5, except that here shown are the results for the model with $M^{\rm II}_{\rm D}$ given in section~3.1.  }
\end{figure*}

In the literature, most of the flavor-symmetry models have been formulated in the MSSM framework for a reason as follows: in order to break the flavor symmetry properly, one needs to introduce some flavon fields which transform as multiplets of the flavor symmetry and develop particular VEV alignments (as we have seen in the above); and the most popular and perhaps natural approach to derive the desired flavon VEV alignments is provided by the so-called F-term alignment mechanism which is realized in the supersymmetric context [6]. For this reason, we will repeat the above study in the MSSM framework.

There are the following three key differences between the leptogenesis in the SM and MSSM frameworks that are relevant for our study: 1) in the MSSM framework one has $y^{2}_\tau = (1+ \tan^2{\beta}) m^2_\tau/v^2$ (with $\tan\beta$ being the ratio of the VEV of the up-type Higgs field to that of the down-type Higgs field), so $\Delta^{}_\tau$ can be greatly enhanced by a large $\tan{\beta}$ value. 2) In the MSSM framework the RGE induced right-handed neutrino mass splitting differs by a factor of 2 from that in the SM framework [i.e., the factor $8\pi^2$ on the right-hand side of Eq.~(26) should be replaced by $4\pi^2$]. 3) In the MSSM framework, in spite of the doubling of the particle spectrum and of the large number of new processes involving superpartners, one does not expect major numerical changes with respect to in the SM framework. To be specific, for given values of $M^{}_I$, $Y^{}_\nu$ and $Y^{}_l$, the total effect of supersymmetry on the final baryon asymmetry can simply be summarized as a constant factor (for a detailed explanation, see section~10.1 of the third reference in Ref.~[13]):
\begin{eqnarray}
\left. \frac{Y^{\rm MSSM}_{\rm B}}{Y^{\rm SM}_{\rm B}} \right|^{}_{M^{}_I, Y^{}_\nu, Y^{}_l} \simeq \left\{ \begin{array}{l} \sqrt{2} \hspace{0.5cm} ({\rm in \ strong \ washout \ regime} ) \; ; \\ 2 \sqrt{2} \hspace{0.5cm} ({\rm in \ weak \ washout \ regime}) \; . \end{array} \right.
\end{eqnarray}

Now we are ready to perform the numerical calculations about the leptogenesis in the supersymmetric context. The numerical results show that the observed value of $Y^{}_{\rm B}$ can be successfully reproduced for sufficiently large values of $\tan \beta$. In Figures~7 and 8 [for the models with $M^{\rm I}_{\rm D}$ and $M^{\rm II}_{\rm D}$ in Eq.~(35), respectively], we have shown the minimal values of $\tan \beta$ that allow for a successful leptogenesis, as functions of the flavor-symmetry scale $\Lambda$. One can see that for large values of $\Lambda$ [which lead to relatively large RGE effects, see Eq.~(51)], one just needs a relatively small value of $\tan \beta$ (around 10) in order to successfully reproduce the observed value of $Y^{}_{\rm B}$. But for smaller values of $\Lambda$ (which lead to relatively small RGE effects), one needs larger values of $\tan \beta$ (a few tens) to enhance the RGE effects so that the observed value of $Y^{}_{\rm B}$ can be successfully reproduced.

\begin{figure}
\centering
\includegraphics[width=6.5in]{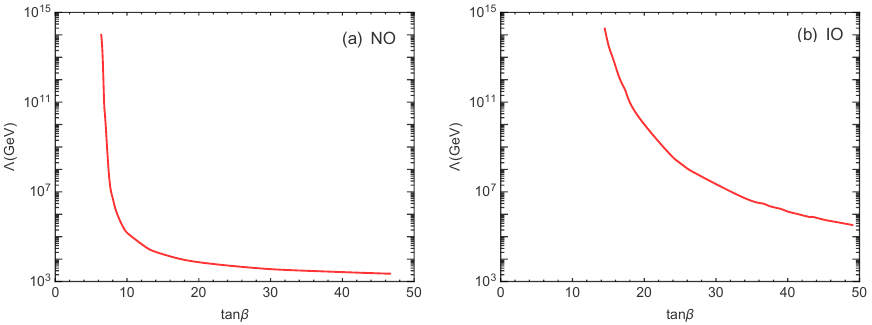}
\caption{ For the model with $M^{\rm I}_{\rm D}$ given in section~3.1, in the MSSM framework, the minimal values of $\tan \beta$ that allow for a successful leptogenesis as functions of the flavor-symmetry scale $\Lambda$, in the NO (a) and IO (b) cases. }
\end{figure}

\begin{figure}
\centering
\includegraphics[width=6.5in]{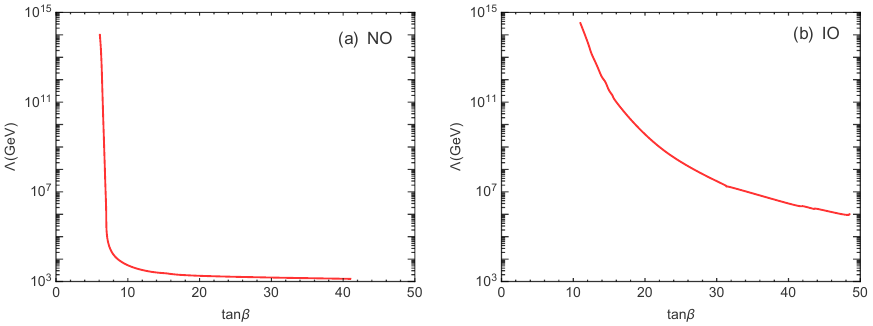}
\caption{ Same as Figure~7, except that these results are for the model with $M^{\rm II}_{\rm D}$ given in section~3.1. }
\end{figure}

Finally, we study what would happen when the $\mu$ term in Eq.~(25) and RGE effects are taken into account simultaneously. In this scenario, the RGE effects will contribute to the breaking of $\mu$-$\tau$ reflection symmetry and the generation of right-handed neutrino mass splitting simultaneously, while the $\mu$ term only contributes to the right-handed neutrino mass splitting. For this scenario, in Figures~9 and 10 [for the models with $M^{\rm I}_{\rm D}$ and $M^{\rm II}_{\rm D}$ in Eq.~(35), respectively], we have shown the maximally allowed values of $Y^{}_{\rm B}$ as functions of the right-handed neutrino mass splitting $\Delta M$. We see that the observed value of $Y^{}_{\rm B}$ cannot be successfully reproduced, except that for the model with $M^{\rm II}_{\rm D}$ in Eq.~(35) there exists a very little parameter space (for $\Delta M \simeq 0.01$ eV) that marginally allow for a successful leptogenesis in the NO case [see Figure~10(a)].

\begin{figure}
\centering
\includegraphics[width=6.5in]{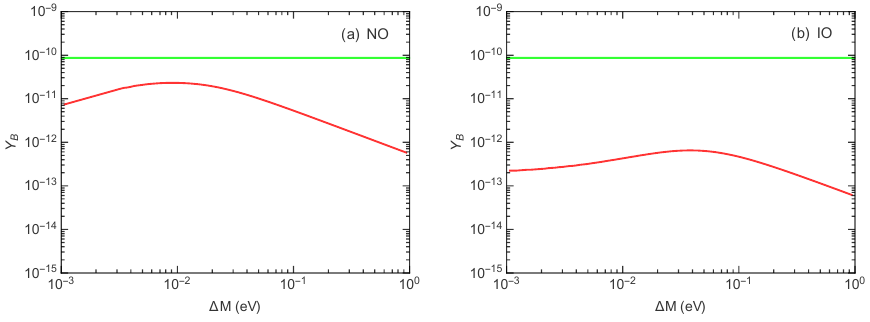}
\caption{ For the model with $M^{\rm I}_{\rm D}$ given in section~3.1, when the $\mu$ term in Eq.~(25) and RGE effects are taken into account simultaneously, the maximally allowed values of $Y^{}_{\rm B}$ as functions of the right-handed neutrino mass splitting $\Delta M$, in the NO (a) and IO (b) cases. The green horizontal line indicates the observed value of $Y^{}_{\rm B}$. These results are obtained by taking $\Lambda =10^{15}$ GeV. }
\end{figure}

\begin{figure}
\centering
\includegraphics[width=6.5in]{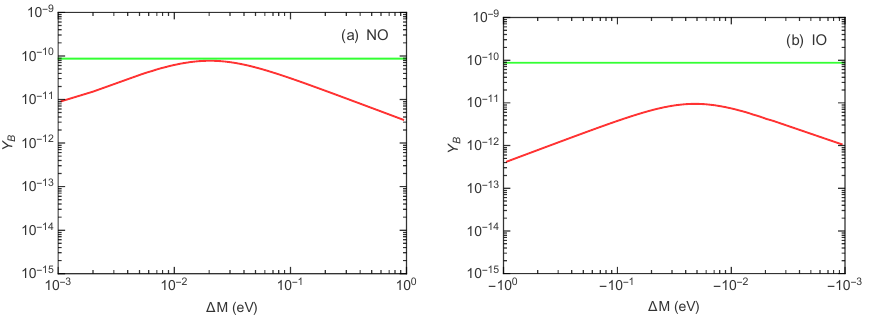}
\caption{  Same as Figure~9, except that these results are for the model with $M^{\rm II}_{\rm D}$ given in section~3.1. }
\end{figure}

\section{Summary}

In this paper, following the Simplicity Principle, we have considered the possibility that there only exists two right-handed neutrinos, and their Majorana mass matrix takes a form as in Eq.~(\ref{9}).
Such a mass matrix can be naturally realized in the minimal linear seesaw model and is the most minimal one in the sense that it only contains a single mass parameter. In this scenario, the two right-handed neutrinos are degenerate in their masses. If they acquire a tiny mass splitting through some way, then the resonant leptogenesis scenario will be naturally realized. In this scenario, a successful leptogenesis can be achieved even if the right-handed neutrino mass $M$ is lowered to the TeV scale, which has the potential to be directly accessed by presently running and forseeable collider experiments.

On the other hand, inspired by the special values of the neutrino mixing angles and a preliminary experimental hint for $\delta \sim -\pi/2$, in the literature the possibility that there may exist a certain flavor symmetry in the lepton sector has been widely studied. The flavor symmetries can serve as a useful guiding principle to help us organize the flavor structure of the neutrino mass model and give some interesting phenomenological consequences. Two popular candidates of them are the TM1 mixing and  $\mu$-$\tau$ reflection symmetry.
Motivated by these facts, we have studied the realizations and consequences of the TM1 mixing and $\mu$-$\tau$ reflection symmetry in the minimal seesaw model with $M^{}_{\rm R}$ in Eq.~(\ref{9}) and TeV-scale right-handed neutrino masses.

In the minimal seesaw model considered in this paper, the TM1 mixing can be naturally realized by taking $M^{}_{\rm D}$ to have a form as shown in Eq.~(\ref{2.1.5}). As discussed below Eq.~(\ref{2.1.5}), such a form of $M^{}_{\rm D}$ can be easily realized by slightly modifying the flavor-symmetry models for realizing the ever-popular TBM mixing. Then, in Figure~1 we have shown the allowed values of the parameters of $M^{}_{\rm D}$ by inputting the neutrino oscillation data. On the basis of these results, we have mainly studied the consequences of this model for leptogenesis. In order for leptogenesis to work, a splitting $\Delta M$ between the two right-handed neutrino masses should arise. In this regard, we have considered two possible ways of generating the tiny splitting between the two right-handed neutrino masses: one way is to modify $M^{}_{\rm R}$ to a form as shown in Eq.~(\ref{2.2.3}) which leads to $\Delta M= 2\mu$; the other way is to consider the renormalization-group corrections for the right-handed neutrino masses which leads to a $\Delta M$ as shown in Eq.~(\ref{2.2.4}). The numerical results show that in both of these two scenarios the observed value of $Y^{}_{\rm B}$ can be reproduced successfully. And Figure~2 has shown the required values of $\mu$ and the flavor-symmetry scale $\Lambda$ for leptogenesis being successful.

On the other hand, the $\mu$-$\tau$ reflection symmetry can be naturally realized by taking $M^{}_{\rm D}$ to have one of the two forms as shown in Eq.~(\ref{3.1.2}). For $M^{\rm I}_{\rm D}$, the values of its parameters can be determined as in Eqs.~(\ref{3.1.8}, \ref{3.1.9}) by inputting the neutrino oscillation data. In this case, in order for leptogenesis to work, one not only needs to generate a splitting between the two right-handed neutrino masses but also needs to break the exact cancellation between the contributions of the $\mu$ and $\tau$ flavors to the baryon asymmetry. The latter can be achieved either with the help of the different conversion efficiencies for the lepton-antilepton asymmetries in different lepton flavors to the baryon asymmetry during the sphaleron processes, or just with the help of the renormalization group evolution effects that serves to generate the splitting between the two right-handed neutrino masses. Unfortunately, the numerical results show that in both of these two scenarios the observed value of $Y^{}_{\rm B}$ cannot be reproduced successfully. Nevertheless, in the supersymmetric SM where the $y^{}_\tau$-related effects can be greatly enhanced by a large $\tan \beta$ value, the observed value of $Y^{}_{\rm B}$ has chance to be reproduced.

For $M^{\rm II}_{\rm D}$, Figure~3 has shown the allowed values of its parameters by inputting the neutrino oscillation data. In this case, due to the vanishing of the CP asymmetries $\varepsilon^{}_{\alpha}$, leptogenesis cannot work unless the $\mu$-$\tau$ reflection symmetry gets broken. Therefore, we have not considered the possibility that the splitting between the two right-handed neutrino masses is realized by modifying $M^{}_{\rm R}$ into the form as shown in Eq.~(\ref{2.2.3}), which keeps the $\mu$-$\tau$ reflection symmetry intact. We have considered the possibility that the splitting between the two right-handed neutrino masses is generated from the renormalization-group corrections, along with which the renormalization group evolution effects also render the breaking of the $\mu$-$\tau$ reflection symmetry. As in the $M^{\rm I}_{\rm D}$ scenario, the renormalization group evolution effects cannot help us successfully reproduce the observed value of $Y^{}_{\rm B}$ in the SM, but can do so in the supersymmetric SM.

\vspace{0.5cm}

\underline{Acknowledgments} \vspace{0.2cm}

This work is supported in part by the National Natural Science Foundation of China under grant Nos.~11605081, 12142507 and 12147214, and the Natural Science Foundation of the Liaoning Scientific Committee under grant NO.~2022-MS-314.

\end{document}